\documentstyle[sprocl]{article}
\def\be{\begin{equation}}
\def\ee{\end{equation}}
\def\bea{\begin{eqnarray}}
\def\eea{\end{eqnarray}}

\begin{document}
\title{ A Study of the Behavior of Ultra-High Energy Neutrinos}
\author{Mou Roy }
\address{Department of Physics,University of California, Riverside ,
 California 92521-0413, USA}
\maketitle\abstracts{
Neutrino Oscillations in the presence of strong gravitational
fields are studied specifically for Majorana neutrinos. We look at
ultra high energy neutrinos $(\sim 1$ PeV) emanating from 
Active Galactic Nuclei (AGN). The spin flavor resonant transitions of
such neutrinos may occur in the vicinity of AGN due to gravitational
effects and due to the presence of a large magnetic field $(\sim 1 $ Tesla).
In this scenario the matter effects (normal MSW transitions) become 
negligible in comparison to gravitational effects. We discuss the
corresponding bound on the magnetic moment (transition magnetic moment
for Majorana neutrinos). Fluxes of different flavor neutrinos are
estimated and probabilities for different neutrino transitions 
calculated.}

\begin{center}
\subsection*{1. Introduction}
\end{center}

Majorana particles are natural representations of massive neutrinos.
Neutrinos in general, and in particular Majorana neutrinos can be used to
probe the core of cosmological objects. Due
to their small cross sections these particles can stream out unaffected
from even the most violent environments such as those in Active
Galactic Nuclei (AGN) \footnote{ AGN are the most luminous objects in
the Universe, believed to be powered by a supermassive black hole of
mass $ 10^4 $ to $ 10^{10} M_\odot$.}. 
In their trek from their source to the
detector the neutrinos can undergo flavor and/or spin transitions which
can obscure some of the features of the source. Because of this, and due
to the recent interest in neutrino astronomy (e.g. DUMAND II, AMANDA,
NESTOR, BAIKAL, etc. \cite{jw}) it becomes important to
understand the manner in which these transitions occur
in the hope of disentangling these effects from the ones produced by the
properties of the source \cite{mou}.

\begin{center}
\subsection*{2. Neutrino Oscillations in AGN environment}
\end{center}

To determine the effective interactions of the Majorana
neutrinos in an AGN environment we start following \cite{mou}, from
the Dirac equation in curved space including their weak and
electromagnetic interactions

\begin{equation}
[i e^\mu_a \gamma^a (\partial_\mu + \omega_\mu) - m + \not\! J \gamma_5
+ \mu \sigma^{ab} F_{ab}] \psi = 0
\label{aaaa}
\end{equation}
where $e^\mu_a$ are the tetrads, $ \gamma^a $ the usual Gamma matrices,
$m$  the mass matrix, $ \not \! J = J_a \gamma^a $ denotes the
weak interaction current matrix, $\mu$ the neutrino magnetic moment,
$F^{ab}$ the electromagnetic field tensor, $ \sigma^{ab} = { 1 \over 4} 
[\gamma_a,\gamma_b ] $;
 and the spin connection is $\omega_\mu={1\over 8}[\gamma_a,\gamma_b] 
e^{\nu a} e^b_{\nu;\mu}$,
where the semicolon denotes a covariant derivative.
We used Greek indices ($\mu, \, \nu , \ldots $) to denote space-time
directions, and Latin indices ($a , \, b , \ldots $) to denote
directions in a local Lorentzian frame.
We have studied in detail the method of extracting the effective 
neutrino Hamiltonian from
eqn.(\ref{aaaa}) in \cite{mou}.
We  have allowed the
possibility of rotation of the central AGN black hole using 
a Kerr metric (we also assume
that the accreting matter generates a small perturbation of the
gravitational field). The metric for a Kerr black hole contains two
parameters, $ r_g $, the horizon radius
and $a$  the total angular momentum of the black hole per unit mass.
Using the typical spherical accretion AGN model \cite{pk}, we can compare 
the magnitude of the weak interaction
current, $J_W \sim 10^{-33} \; \rho \;\; {\rm eV}^{-1}$, to its 
gravitational counterpart $J_G \sim r_g^{-1}$.
The order of matter density $\rho$ for typical cases is $10^1-10^4 \;{\rm eV
}^4$ which shows that
the  gravitational current part dominates
the weak current part for all relevant values of $ r_g  \; (10^{14}\; {\rm to}
\; 10^{20} {\rm eV^{-1}} )$. This  causes the normal MSW effects to be 
negligible
in this scenario.

\begin{center}
\subsection*{3. Probabilities of Allowed Transitions and Neutrino Flux
Modifications.}
\end{center}

In the adiabatic limit it is easy to quantitatively describe the
gravitational effect in AGN environment
by looking at the transition and survival probabilities~\footnote{The 
procedure is
essentially the same as the one used in describing the MSW
resonances.}.
The condition for resonances to
induce an appreciable transition probability 
gives
a bound on the neutrino magnetic moment $\mu$ (transition magnetic moment).
For a study  of the probabilities it is reasonable
to choose a specific value of $\mu$ which would ensure neutrino transitions
for all the energy and  $ \Delta m^2 $ \footnote{ $ \Delta m^2 = m^2_1 -
m^2_2 $ } values 
of interest. We have chosen $ \mu_t = 10^{-13} \mu_B $~\footnote{
For $ E=1 $TeV and $ \Delta m^2 = 10^{-6} $eV$^2$,
$ \mu^{\rm res}_{\rm min} \sim 10^{-13} - 10^{-14} \mu_B $.}.
The probabilities
for a specific energy and $ \Delta m^2 $ remains approximately
constant with $(r,\theta)$. However
transition probability increases with energy and decreases with
$\Delta m^2 $ (reverse behavior for the survival probability).
For simplicity we have limited ourselves to only two neutrino mixing.

The high energy neutrino detectors
are sensitive to a wide range of
neutrino parameters and will be able to test a variety of
models of neutrino production in AGN \cite{pakvasa}
for neutrino energies over 1 TeV.
Matter effects are negligible in the AGN
environment,
but gravity-induced resonances could cause
a decrease of the neutrino flux of any given flavor independent of the
value of $\Delta m^2$ chosen. This effect
could cause an oscillation to $\tau $ neutrinos \footnote{ There is
negligible $\tau $ neutrino production in the AGN environment according
to all standard models .} generating a significant flux of $ \tau $
neutrinos to which planned experiments will be sensitive in the
PeV range \cite{pakvasa}.
Around this energy
we find that the muon neutrino flux decreases by about 50\% for the
range of values of $\Delta m^2 $ considered. This effect would
decrease the number of
upward moving muons beyond energy 1 PeV by a factor of 2.
However if we consider oscillation between muon and electron neutrinos
the flux reduction is about 25\% at this energy
\footnote{ This is due to the fact
that unlike tau neutrinos, the initial $ ( \nu_e + {\bar \nu}_e) $ flux in AGN 
is not negligible,
but is half of the initial $( \nu_\mu + {\bar \nu}_\mu) $ flux.}.

\begin{center}
\subsection*{5. Conclusions}
\end{center}
We found that ultra high energy Majorana neutrinos will be strongly 
affected by
gravitational and electromagnetic effects provided $ \Delta m^2 >
10^{-10} $eV$^2$ and that the transition magnetic moment satisfies $
\mu_t > 10^{-13} \mu_B $.
Gravitational oscillations are the dominant mechanism for
the flavor/spin oscillation of such neutrinos emanating from 
AGN causing a significant muon neutrino decrease 
and increase of the corresponding $ \tau $ neutrino flux,
to which future neutrino telescopes \cite{pakvasa} will be sensitive.
In the case of solar
neutrinos the presence of more than two flavors can significantly alter the
predicted fluxes. However for the present situation where
the experimental information on the AGN neutrino flux is
quite limited, it is sufficient to determine the various effects and
their strengths by using 
a two flavor mixing description as we have done in our analysis.

\section*{Acknowledgments}  

I would like to thank Prof. Sandip Pakvasa for useful suggestions.
This work was supported in part by the US Department of Energy
under contract FDP-FG03-94ER40837.


\begin{thebibliography}{99} 

\bibitem{jw}
R.J.Wilkes, in {\sl Proc.Slac.Summer
Institute 1994} edited by Jennifer Chan and Lilian De Porcel;
H.W.Sobel, Nucl.Phys.{\bf B} (Proc Suppl) {\bf 19}, 444 (1991);
S.Barwick {\it et.al.}, J.Phys.G: Nucl.Part.Phys.{\bf18}, 225 (1992).

\bibitem{mou} 
D.Piriz, M.Roy and J.Wudka, Phys.Rev.{\bf D54}, 1587 (1996);
M.Roy and J.Wudka, UCRHEP-T174, submitted to Phys.Rev.{\bf D}.

\bibitem{pk}
 R.J.Protheroe and D.Kazanas, Ap.J.{\bf265}, 620 (1983);
D.Kazanas and D.C.Ellison, Ap.J.{\bf304}, 178 (1986).

\bibitem{pakvasa}
J.G.Learned and S.Pakvasa, Astropart. Physics {\bf 3}, 267 (1995); 
R.Gandhi, {\it et.al.}, hep-ph/9604276 (unpublished), report number 
AZPH-TH-96-12.

\end{thebibliography}
\end{document}